\documentclass[a4paper]{jpconf}
\usepackage{graphicx}
\begin{document}
\title{Pressure-induced new magnetic phase in Tl(Cu$_{0.985}$Mg$_{0.015}$)Cl$_3$ probed by muon spin rotation}

\author{Takao Suzuki$^1$, Isao Watanabe$^1$, Fumiko Yamada$^2$, Motoki Yamada$^2$, Yasuyuki Ishii$^1$, Takayuki Kawamata$^1$, Takayuki Goto$^3$ and Hidekazu~Tanaka$^2$}

\address{$^1$RIKEN Nishina Center for Accelerator-Based Science, 2-1 Hirosawa, Wako, Saitama 351-0198, Japan}

\address{$^2$Department of Physics, Tokyo Institute of Technology, O-okayama, Meguro-ku, Tokyo 152-8551, Japan}

\address{$^3$Faculty of Science and Technology, Sophia University, 7-1 Kioi-cho, Chiyoda-ku, Tokyo 102-8554, Japan}

\ead{suzuki\_takao@riken.jp}

\begin{abstract}
%\indent
We carried out zero-field muon-spin-rotation (ZF-$\mu$SR) measurements in hydrostatic pressures in impurity-doped quantum spin gap system Tl(Cu$_{0.985}$Mg$_{0.015}$)Cl$_3$ to investigate microscopic magnetic properties of the pressure-induced phase.
The spontaneous muon spin precession, which indicates the existence of a long-range coherent order, is observed in pressures. With decreasing temperature in 3.1 kbar, the internal static magnetic field at the muon sites {\it H}$_{\rm int}$ tends to saturate to 280 Oe around 4 K, however, decreases to 240 Oe at 2.3 K.
These results suggest a rearrangement of ordered spins, and we speculate that the oblique antiferromagnetic phase, which is observed in the pressure of 14 kbar on the pure TlCuCl$_3$, appears in the Mg-doped system in lower pressures.
\end{abstract}

\section{INTRODUCTION}
\indent\indent
TlCuCl$_3$ is the parent material of the subject compound in this study, and is three-dimensionally coupled Cu-3{\it d} S = 1/2 spin dimer system.
This system has the magnetic ground state of spin-singlet with excitation gap of $\Delta_{\rm gap} =$ 7.5 K, which originates from strong intradimer antiferromagnetic interaction~\cite{Shiramura,Cavadini}.
In the impurity-introduced Tl(Cu$_{1-x}$Mg$_x$)Cl$_3$ system, the magnetic phase transition to an ordered state is observed by magnetization and specific heat measurements, and neutron elastic scattering measurements identified that this impurity-induced ordered state is the antiferromagnetically ordered state of which the magnetic structure is the same with the case of the field-induced phase in TlCuCl$_3$~\cite{Oosawa_Mg,Fujisawa_Mg}.
The inelastic neutron scattering measurement revealed that a finite spin gap still remains below the transition temperature in Tl(Cu$_{1-x}$Mg$_x$)Cl$_3$~\cite{Oosawa_Mg2}.
Imamura {\it et al}. reported the pressure-induced magnetically ordered phase by magnetization measurements in Tl(Cu$_{1-x}$Mg$_x$)Cl$_3$ with $x =$ 0.012, and concluded that the change from the impurity-induced phase to the pressure-induced phase is the crossover~\cite{Imamura_Mg}.
In the case for the slightly doping of $x =$ 0.0047, longitudinal-field muon-spin-relaxation measurements revealed that the impurity-induced magnetic moments of the Cu-3{\it d} spins slowly fluctuate at $T =$ 20 mK, far below the magnetic phase transition temperature observed by the specific heat measurement~\cite{Suzuki_f}.
Recently, Yamada {\it et al}. reported the spin reorientation phase transition at {\it T}$_{\rm R} =$ 9.2 K in the high pressure of 14 kbar in the pure TlCuCl$_3$, and discussed a second-order transition from the spin-flop phase to the oblique antiferromagnetic phase below {\it T}$_{\rm R}$\cite{Fumiko}.
Magnetic orderings induced by impurity doping and by pressure in spin gap systems are still attracting much interest.
And besides, it has been revealed that muons are the excellent probe to detect magnetic properties of randomness-introduced spin gap systems~\cite{Saito,Goto,Suzukix02,Suzukix06}.
The purpose of this study is to investigate the microscopic properties of the pressure-induced phase in the impurity-doped spin gap system by the zero-field muon-spin-rotation/relaxation (ZF-$\mu$SR).

\section{EXPERIMENTS}
\indent\indent
Single crystals used in this study were grown from a melt by the Bridgman method.
The details of crystal growth are given elsewhere~\cite{Oosawa_TlK}.
The concentration of $x$ was determined by the inductively coupled plasma atomic emission spectrometry (ICP-AES) method.
Muon-spin-rotation/relaxation ($\mu$SR) measurements were performed on Tl(Cu$_{1-x}$Mg$_x$)Cl$_3$ with $x =$ 0.015 at the RIKEN-RAL Muon Facility~\cite{Matsuzaki}.
Measurements in hydrostatic pressures were carried out using a spin-polarized double-pulsed positive decay-muon beam with an incident muon momentum of 90 MeV/c.
Samples are pressurized by $^4$He gas pressure in a CuBe cell using the newly installed gas-pressurized $\mu$SR setup for the RIKEN-RAL Muon Facility~\cite{Nabedon}.
Cleaved crystals were packed into the pressure cell in random directions.
$\mu$SR measurements were carried out at two separate beam times.
Single crystals were cleaved from the same rod for each beam time.
In $\mu$SR measurements, spin-polarized muons are implanted into samples.
The asymmetry was defined as follows:
\begin{eqnarray}
A(t)=\frac{F(t)-\alpha B(t)}{F(t)+\alpha B(t)}
\end{eqnarray}
Here, $F(t)$ and $B(t)$ were total muon events counted by the forward and backward counters at time $t$ respectively.
The $\alpha$ is a calibration factor reflecting relative counting efficiencies between the forward and backward counters.
The initial asymmetry is defined as $A(0)$.
Measured time spectra were analyzed using the WiMDA computer program~\cite{Wimda}.

\section{RESULTS AND DISCUSSIONS}
\indent\indent
Figure 1 shows the temperature dependence of ZF-$\mu$SR time spectra in hydrostatic pressure of 3.1 kbar.
(The pressure dependence of time spectra at 2.3 K is reported in elsewhere~\cite{Suzuki_Mg}.)
Measurements were carried out with decreasing temperature.
Below 8 K, the spontaneous muon-spin-precession is observed, which indicates the existence of a coherent long-range magnetically ordered state.
The observed transition temperature in pressure is consistent with the reported temperature in the case of $x =$ 0.012 deduced from the magnetization measurement\cite{Imamura_Mg}.
In order to discuss the development of the internal static magnetic field at the muon sites which corresponds to the ordered Cu-3{\it d} magnetic moment, all spectra are analyzed using the two components function as follows:
\begin{eqnarray}
A(t)=A_1e^{-\lambda_1 t}\cos(\omega t + \theta) + A_2e^{-\lambda_2 t}G_{\rm z}(\Delta,t)
\end{eqnarray}
The first term is the signal from the magnetically ordered region of samples, and the second term is that from the large pressure cell and the spin fluctuating region of samples, because it is quite difficult to distinguish the relaxing part $\exp (-\lambda t)$ between the pressure cell and the samples.
$\lambda_1$ and $\lambda_2$ are the muon-spin-relaxation rate of each component, and $\omega$ is the muon-spin-rotation frequency.
$G_{\rm z}(\Delta,t)$ is the static Kubo-Toyabe function, where  $\Delta/\gamma_{\mu}$ is the distribution width of nuclear-dipole fields at the muon sites.
$\gamma_{\mu}$ is the gyromagnetic ratio of the muon spin ($2\pi\times 13.5534$ ${\rm kHz/gauss} $).
The ratio of the amplitude $A_1/A_2$ is approximately 0.16 above 4 K.
Fitted results are shown in Fig.1 as solid lines.
From the fitted result, the internal static magnetic field $H_{\rm int}$ at the muon sites is deduced using the relation of $\omega = \gamma_{\mu}H_{\rm int}$ because implanted muon spins precess around the total magnetic field of the internal field and the external field at the muon sites.

\begin{figure}[ttt]
\vspace*{-1.5cm}
\begin{minipage}{14pc}
\begin{flushleft}
\includegraphics[width=20pc]{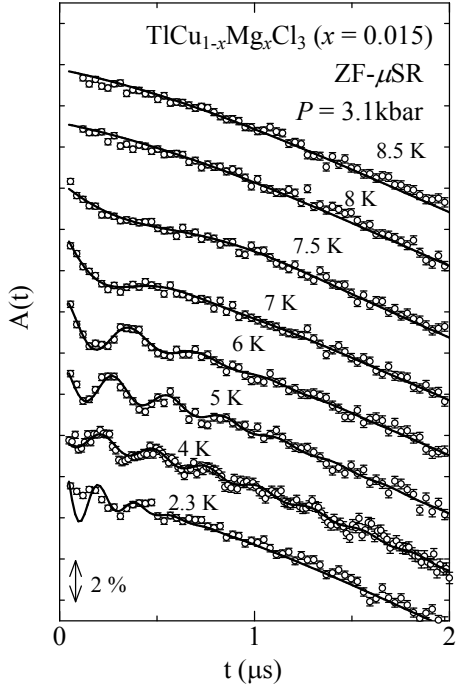}
\end{flushleft}
\end{minipage}\hspace{6pc}%
\begin{minipage}[t]{18pc}\caption{\label{label}Time spectra of the ZF-$\mu$SR in TlCu$_{1-x}$Mg$_{x}$Cl$_3$ with $x$ = 0.015 in the hydrostatic pressure of 3.1 kbar in the temperature range from 8.5 K to 2.3 K.
Solid lines are fitted results using the function (2).
Each plot is shifted consecutively for clarity.}
\end{minipage}
\vspace*{-1.6cm}
\end{figure}

\begin{figure}[ttt]
\vspace*{-2.6cm}
\begin{minipage}{14pc}
\begin{flushleft}
\includegraphics[width=22pc]{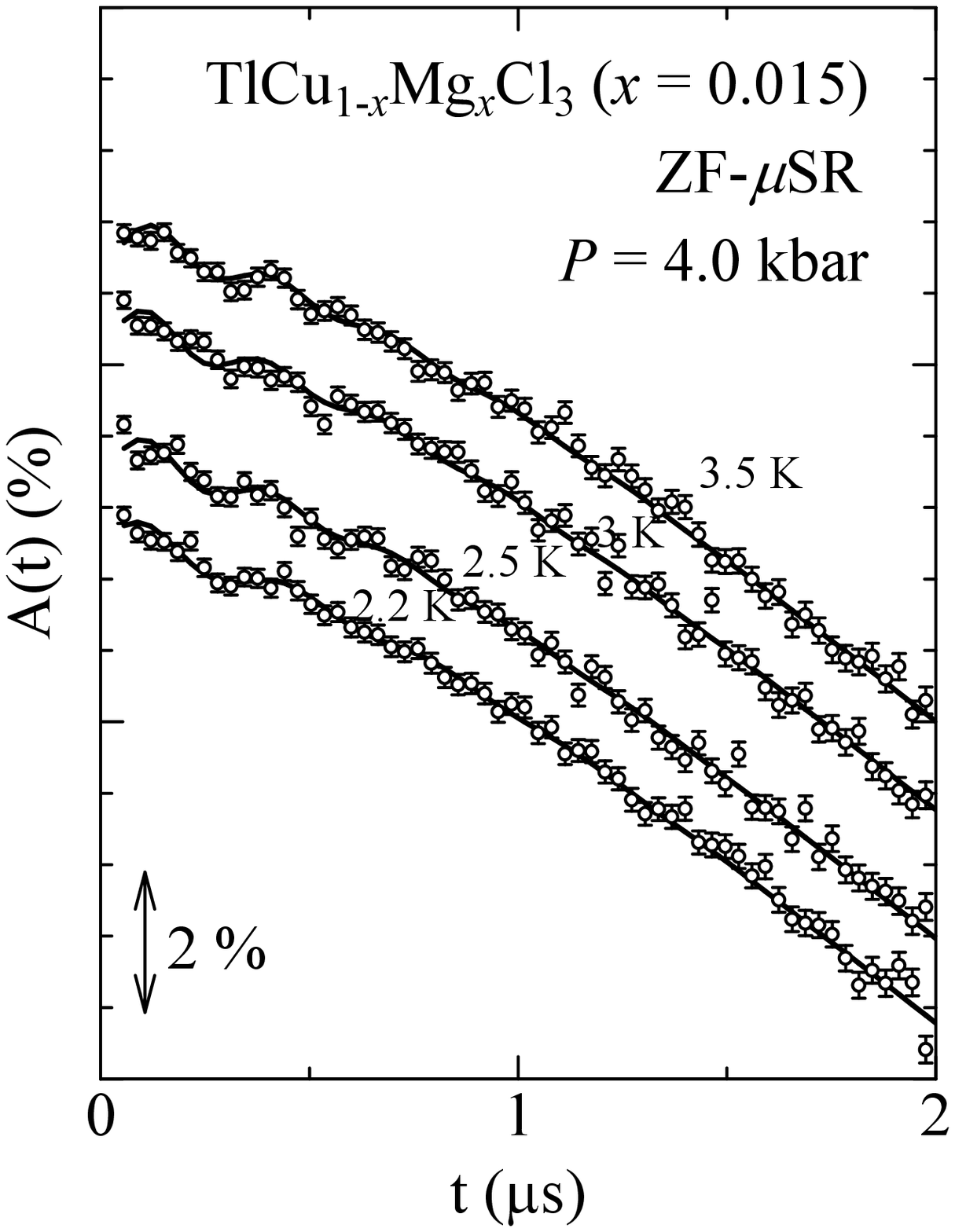}
\end{flushleft}
\end{minipage}\hspace{6pc}%
\begin{minipage}[t]{18pc}\caption{\label{label}Time spectra of the ZF-$\mu$SR in TlCu$_{1-x}$Mg$_{x}$Cl$_3$ with $x$ = 0.015 in the hydrostatic pressure of 4.0 kbar.
Solid lines are fitted results using the function (2).
Each plot is shifted consecutively for clarity.}
\end{minipage}
\vspace*{-2.8cm}
\end{figure}

\begin{figure}[tbh]
\vspace*{-4.0cm}
\begin{minipage}{14pc}
\begin{flushleft}
\includegraphics[width=26pc]{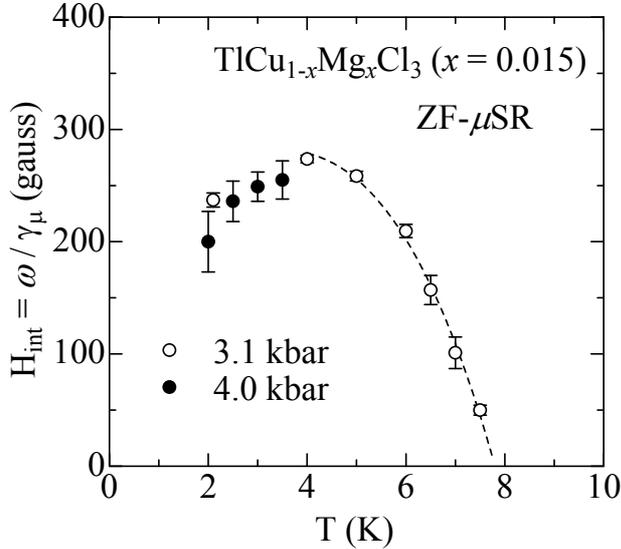}
\end{flushleft}
\end{minipage}\hspace{8pc}%
\begin{minipage}[t]{16pc}\caption{\label{label}Temperature dependence of the internal static magnetic field $H_{\rm int}$ in 3.1 kbar (open circles) and in 4.0 kbar (closed circles).
Dashed line is a guide for the eye.}
\end{minipage}
\vspace*{-4.2cm}
\end{figure}

Figure 2 shows the temperature dependence of ZF-$\mu$SR time spectra in hydrostatic pressure of 4.0 kbar.
Spectra are analyzed in the same way with in the case for 3.1 kbar, but the ratio of the amplitude $A_1/A_2$ at 3.5 K is 0.04 in the case of 4.0 kbar.
The small amplitude of the rotational signal is partially due to a slight discrepancy in the momentum tuning for implanting muons in samples and is partially due to the decrease of the volume fraction of the coherently ordered region at lower temperatures as discussed below.
Fitted results are shown as solid lines.
Deduced internal static magnetic field $H_{\rm int} = \omega / \gamma_{\mu}$ at the muon sites are summarized in Fig.3.
With decreasing temperature in 3.1 kbar, $H_{\rm int}$ increases monotonously, and tends to saturate to 280 Oe around 4 K.
However, $H_{\rm int}$ decreases to 240 Oe at 2.3 K.
Similar decrease of $H_{\rm int}$ below 4 K is observed in 4.0 kbar.
\\
\indent
The decrease of $H_{\rm int}$ suggests an orientation of the ordered moments, because the change of the magnetic structure leads to the modification of dipole fields at the muon sites.
As mentioned above, in the pure TlCuCl$_3$, the spin reorientation phase transition occurs at {\it T}$_{\rm R} =$ 9.2 K in the high pressure of 14 kbar.
In the pure sample, the spin gap is collapsed by pressure in $P_{\rm c} =$ 0.42 kbar~\cite{Kgoto}.
For the Mg-doped case, a finite spin gap still remains although the magnetic phase transition occurs~\cite{Oosawa_Mg2}, and it is expected that the pressure effect is more significant in the case for Mg-doped system, because the spin gap has already partially collapsed in space by the Mg-doping.
In other word, the pressure-induce new phase could be appeared in lower pressures compared to the case in the pure system.
Thus, we speculate that the spin reorientation phase transition occurs and that the oblique antiferromagnetic phase, which is observed in the pressure of 14 kbar on the pure TlCuCl$_3$~\cite{Fumiko}, appears in the Mg-doped system.
\\
\indent
In the last, we discuss the decrease of the rotational amplitude in time spectra below 4 K.
Generally, the amplitude of the rotational signal corresponds to the volume fraction of the coherent long-range ordered region.
For that reason, the decrease of the rotational amplitude suggests that, after the spin reorientation phase transition, the spin system become less coherent due to some randomness caused by the spin reorientation.
In other word, some randomness caused by the spin reorientation would disturb the coherency of the ordered state.

\section{SUMMARY}
\indent\indent
Zero-field muon-spin-rotation (ZF-$\mu$SR) measurements in hydrostatic pressures were carried out in impurity-doped quantum spin gap system Tl(Cu$_{1-x}$Mg$_x$)Cl$_3$ with $x =$ 0.015.
The spontaneous muon spin precession, which indicates the existence of a long-range coherent order, is observed under pressures.
With decreasing temperature in 3.1 kbar, the internal static magnetic field at the muon sites $H_{\rm int}$ tends to saturate to 280 Oe around 4 K, however,  decreases to 240 Oe at 2.3 K.
Similar decrease of $H_{\rm int}$ below 4 K is observed in 4.0 kbar.
These results suggest a rearrangement of ordered spins which leads to the change of $H_{\rm int}$.
We speculate that the oblique antiferromagnetic phase, which is observed in the pressure of 14 kbar on the pure TlCuCl$_3$, appears in the Mg-doped system.

\section*{Acknowledgement}
\indent\indent
We would like to thank C. M. Goodway and all other members of high-pressure group of ISIS Facility at the Rutherford Appleton Laboratory for the technical support on $\mu$SR measurements in pressures.

\end{document}